\documentclass[prb,preprint,showpacs]{revtex4} %% single column, preprint

\usepackage{graphicx}
\usepackage{hyperref}
\usepackage{amsmath}

\begin{document}

\title{Statistical studies of random lasing modes and amplified spontaneous emission spikes in weakly scattering systems}

\author{X. Wu and H. Cao}

\affiliation{Department of Physics and Astronomy, Northwestern University, Evanston, Illinois, 60208\\}

\begin{abstract}
We measured the ensemble-averaged spectral correlation functions and statistical distributions of spectral spacing and intensity for lasing modes in weakly scattering systems, and compared them to those of the amplified spontaneous emission spikes.  Their dramatic differences illustrated the distinct physical mechanisms. Our numerical simulation revealed that even without reabsorption the number of potential lasing modes might be greatly reduced by local excitation of a weakly scattering system. The lasing modes could be drastically different from the quasimodes of the passive system due to selective amplification of the feedback from the scatterers within the local gain region.   
\end{abstract} 

\maketitle

\section{Introduction}

The random laser differs from the conventional laser in that the optical feedback originates from light scattering instead of reflection \cite{cao_WRM}. One important topic of research is the nature of random laser modes. In his seminal paper \cite{letokhov}, Letokhov predicted lasing with non-resonant feedback in a diffusive random medium. The lasing modes are the eigenmodes of the diffusion equation, and the lasing frequency is equal to the center frequency of the amplification line. In spite of drastic narrowing of emission spectrum at the threshold, the lasing spectrum above the threshold remained relatively broad and featureless \cite{lawandy}. In contrast, multiple narrow peaks were observed on top of the amplified spontaneous emission spectrum in strongly scattering semiconductor powders and polycrystalline films \cite{caoAPL98,caoPRL99}. Frequencies of the discrete lasing peaks, which cannot be selected by the smooth gain spectrum, depend on the spatial distribution of the dielectric constant. Namely, they change with the spatial configuration of the scatterers, but remain constant for a fixed configuration. Simulation and experiment showed that for a random system in the localization regime lasing oscillation occurs in the eigenmodes of Maxwell's equations (also called quasimodes for an open system) \cite{vanneste_2001,jiang_2002,genack_2005}. The interference of scattered light plays a key role in the formation of quasimodes or lasing modes. After multiple scattering, light (of wavelength $\lambda$) may return to a coherence volume ($\sim \lambda^3$) it has visited before, providing field feedback for lasing. The lasing frequencies are determined by the interference of the scattered light returning via different {\it closed} paths. Thus the feedback is resonant (frequency-dependent) and coherent (phase-sensitive). Surprisingly, random lasing with resonant feedback was also realized for various systems in the diffusive or even ballistic regime, despite the coherent interference effect is expected to be negligibly small \cite{frolov_1999,ling_2001,polson_2005,wu_2006}. The lasing modes are believed to be the quasimodes of the passive random system, in particular, the ones with small decay rate \cite{apalkov_2002,cao_2002,vanneste_2007}. The statistics of the decay rates of the quasimodes has been used to predict the lasing threshold and the number of lasing modes \cite{burin_2001,apalkov_2002,patra_2003,hacken_2005,pinheiro_2006,lagendijk_2006}. However, recent theoretical studies suggest that the quasimodes of a passive random system may not be the genuine normal modes of the same system with gain \cite{deych_2005,tureci_2006,tureci_2007}. A careful investigation of the lasing modes in weakly scattering systems and their relation to the quasimodes of the passive systems is needed to understand the random lasing phenomena.

The spatial inhomogeneity of optical gain makes the situation even more complicated. Experimentally, tight focusing of the pump beam is necessary to observe discrete lasing peaks. The number of lasing peaks increases with the pumped spot size, and eventually they merge to a continuous broad band. This phenomenon can be easily explained for the random systems in or near the localization regime. A large pumped volume contains many localized modes which may serve as lasing modes under excitation. Some of the lasing modes are spatially separated and have similar frequencies. However, in diffusive or ballistic systems, the quasimodes are spatially extended across the entire system, except the rare anomalously localized modes \cite{apalkov_2002}. Even with local pumping, all modes of frequency within the gain spectrum are excited almost equally. Therefore shrinking the gain volume should not decrease the number of {\it potential} lasing modes, but only increase their lasing threshold. This seemingly contradiction originates from the assumption that the quasimodes of the passive system serve as the lasing modes. Our recent numerical and experimental studies demonstrate that this assumption no longer holds when reabsorption of emitted light outside the pumped region is strong \cite{yamilov_2005,wu_2006}. The suppression of the feedback from the unpumped part effectively reduces the system size and the lasing modes are confined in the vicinity of the pumped region.  The number of quasimodes of the ``reduced'' system is therefore much less than that of the total system, facilitating the observation of discrete lasing peaks. However, this theory is not applicable to the systems with negligible reabsorption outside the excitation region. The effect of local pumping on random lasing modes in a weakly scattering system without reabsorption is not fully understood. 

In addition to the lasing peaks, stochastic spikes were reported in the single-shot spectra of amplified spontaneous emission (ASE) from random systems over a wide range of scattering strength \cite{mujumdar_2004}. The spikes are intrinsically stochastic and vary from shot to shot even though the disorder is static \cite{mujumdar_2007}. They are attributed to single spontaneous emission events which happen to take long {\it open} paths inside the amplifying random medium and pick up large gain. The emergence of ASE spikes does not rely on the resonant feedback or the coherent interference. One question that arises in the experiment is how to distinguish the ASE spikes from the lasing peaks, especially for the colloidal solutions in which the particles move constantly and the lasing peaks also change from shot to shot. Our recent experimental data illustrate that the statistics of lasing peaks is distinct from that of ASE spikes \cite{wu_2007a}. However, the physical origina of the statistical behavior of random lasing modes is still not clear.     

The above questions and issues will be addressed in this paper. First we will present a detailed experimental study to illustrate the fundamental difference between the ASE spikes and the lasing peaks. The ASE spikes can appear in the absence of scattering, while the lasing peaks rely on the coherent feedback provided by scattering. The ensemble-averaged spectral correlation functions for the ASE spikes and lasing peaks, as well as the statistical distributions of their spectral spacings and intensities, are drastically different. Such differences underline their distinct physical mechanisms. Next we investigate the origin of coherent random lasing in weakly scattering systems and the effect of local pumping on the lasing modes in the absence of reabsorption. Despite of the weak scattering, the coherent feedback may be strongly enhanced due to a large amplification. Our numerical simulations reveal that local excitation can greatly reduce the number of lasing modes even without reabsorption outside the pumped region. It increases the frequency spacing of lasing modes, and facilitates the observation of discrete peaks in spectrum. The lasing modes, though still extended across the entire system, may differ dramatically from the quasimodes due to selective amplification of the feedback from the scatterers within the local gain region.

\section{Lasing peaks vs. ASE spikes}

Our experiments were performed on the diethylene glycol solutions of stilbene 420 dye and TiO$_2$ particles (mean radius = 200nm). The colloidal solution was first shaken in an ultrasonic bath for 30 min and then transfered to a quartz cuvette of dimension $1 \times 1 \times 5$cm. The solution was excited by the third harmonics of a pulsed Nd:YAG laser at wavelength $\lambda_p=355$nm. The pulse width was 25ps and the repetition rate was 10Hz. The pump beam was focused by a lens into the solution through the front window of the cuvette. The spot of pump beam at the front window was roughly 30$\mu{m}$ in diameter. The emitted light was collected by the same lens, then focused by a second lens to a fiber bundle which was connected to a spectrometer with a cooled CCD array detector. 

The reason for choosing stilbene 420 instead of the commonly used rhodamine dye is the weak reabsorption outside the pumped volume. Since the absorption band of stilbene 420 is well separated from its emission band, the absorption at the emission wavelength is extremely weak. For example, the absorption length $l_a$ at the center emission wavelength $\lambda_e$ = 427nm was 6cm at the dye concentration $M$ = 8.5mM. It was much larger than the dimension $L$($\sim$ 1cm) of the cuvette that held the solution. At the particle density $\rho = 3\times10^{9}$cm$^{-3}$, the scattering mean free path $l_s \simeq 1.3$mm at $\lambda_p$, and $l_s \simeq 1.0$mm at $\lambda_e$. Although $l_{a} \simeq 10\mu$m at $\lambda_p$, the pump light penetrated much deeper than $l_a$ due to the saturation of absorption by intense pumping. Imaging through the side window of the cuvette revealed that the excitation volume had a cone shape of length a few hundred micron and base diameter 30$\mu$m. Because the cone length was smaller than $l_s$, the excitation cone in the colloidal solution was almost identical to that in the neat dye solution ($\rho =0$). The transport of emitted light was diffusive in the colloidal solution with $\rho = 3\times10^{9}$cm$^{-3}$, because the dimension $L$ of the entire solution was much larger than $l_s$. Light amplification, however, occurred only in a sub-mean-free-path volume which was pumped. The motion of the particles in the solution provided different random configuration for each pump pulse, which facilitated the ensemble measurement under identical conditions.
  
The single-shot emission spectra from the colloidal solution with $\rho$ = $3\times 10^9$cm$^{-3}$ and $M$ = 8.5mM are shown in Figs. 1(a) to (c) with increasing pump pulse energy $E_p$. At $E_p$ = 0.05$\mu$J [Fig. 1(a)], the spectrum exhibited sharp spikes on top of a broad ASE band.  From shot to shot the spikes changed completely. The typical linewidth of the spikes was about 0.07nm. The neighboring spikes often overlapped partially. As pumping increased, the spikes grew in intensity. When $E_p$ exceeded a threshold, a different type of peaks emerged in the emission spectrum as shown in Fig. 1(b). They grew much more rapidly with pumping than the spikes, and dominated the emission spectrum at $E_p$ = 0.13$\mu$J [Fig. 1(c)]. The peaks, with  the typical linewidth of 0.13nm, were notably broader than the spikes. Unlike the spikes, the spectral spacing of adjacent peaks was more or less regular. We repeated the experiment with solutions of different $\rho$ as well as the neat dye solution of the same $M$. The peaks could only be observed with particles in the solution, while the spikes appeared also in the spectrum of emission from the neat dye solution. Figures 1(d)-(f) are the emission spectra of the neat dye solution with the same pump pulse energies as those of the colloidal solution in Figs. 1(a)-(c). Although they were similar at $E_p$ = 0.05$\mu$J, the emission spectra of these two samples were dramatically different at $E_p$ = 0.13$\mu$J. Even under intense pumping, the emission spectrum of the neat dye solution had only spikes but no peaks [Fig. 1(f)]. The maximum spike intensity was about 50 times lower than the maximum peak intensity in the colloidal solution at the same pumping [Fig. 1(c)]. The pump threshold for the appearance of peaks depended on the particle density in the solution, while the threshold for the emergence of spikes in the solutions of low $\rho$ was similar to that in the neat dye solution. 

In our previous experimental and numerical studies \cite{wu_2006}, we concluded that the large peaks represented the lasing modes formed by distributed feedback in the colloidal solutions. Although the feedback was weak at low particle density, the intense pumping strongly amplified the backscattered light and greatly enhanced the feedback. In contrast, the feedback from the particles was not necessary for the spikes which also existed in the neat dye solution.  Thus the spikes were attributed to the amplified spontaneous emission. To demonstrate quantitatively the difference between the ASE spikes and the lasing peaks, we investigated their spectral correlations and intensity statistics. Since it was difficult to obtain reliable statistical data for the ASE spikes from the colloidal solution at high pumping due to the presence of dominant lasing peaks, we obtained the data of ASE spikes from the neat dye solution instead.

\section{Spectral correlations}

The ensemble-averaged spectral correlation function $C(\Delta\lambda)$ was obtained from 200 single-shot emission spectra acquired under constant pumping condition. We chose the wavelength range 425-431nm, within which the gain coefficient showed very small change, to compute $C(\Delta\lambda)= \langle I(\lambda )I(\lambda+\Delta\lambda) \rangle / \langle I(\lambda)\rangle \langle I(\lambda+\Delta\lambda) \rangle -1$. Figure 2 shows $C(\Delta\lambda)$ obtained with the colloidal solution and the neat dye solution at various pumping levels. For the colloidal solution ($\rho$ = $3\times 10^9$cm$^{-3}$, $M$ = 8.5mM), $C(\Delta\lambda)$ changed dramatically with increasing pumping [Fig. 2(a)]. Below the lasing threshold where the emission spectrum had only ASE spikes, the value of $C(\Delta \lambda)$ was small at $\Delta \lambda = 0$, and it decayed quickly to zero as $\Delta \lambda$ increased. Once the pumping exceeded the threshold and lasing peaks emerged, the amplitude of $C(\Delta\lambda)$ grew rapidly, and regular oscillations with $\Delta\lambda$ were developed. The oscillation period was about 0.27nm, corresponding to the average wavelength spacing of adjacent lasing peaks. Due to the slight variation of the lasing peak spacing, the oscillation was damped and the correlation peaks were broadened with increasing $\Delta\lambda$. Nevertheless, the oscillation of $C(\Delta \lambda)$ survived the ensemble average despite the lasing peaks changed from shot to shot. This result confirmed not only the lasing peaks in a single-shot spectrum were more or less regularly spaced, but also the average peak spacing was almost the same for different shots. In contrast, $C(\Delta \lambda)$ for the neat dye solution barely changed with increasing pumping once the ASE spikes emerged [Fig. 2(b)]. Although the amplitudes of ASE spikes increased with pumping, their spectral correlations remained almost the same. $C(\Delta \lambda)$ for the neat dye solution was similar to that of the colloidal solution below the lasing threshold where the spectrum had only ASE spikes. Note that the ASE spikes produced irregular oscillations in the spectral correlation function of a single shot emission spectrum. However, such oscillations were removed after averaging over many shots. This result reflected the stochastic nature of the ASE spikes.

We also obtained the statistical distribution $P(\delta \lambda)$ of wavelength spacing $\delta \lambda$ of adjacent lasing peaks and that of the ASE spikes. There were several ways of identifying the peaks/spikes in the emission spectra, e.g., three-point peak-finding, five-point peak picking. The spectral resolution of our spectrometer was limited by the pixel size of the CCD array detector. Each pixel corresponded to a wavelength interval $d \lambda$ = 0.02nm. In an emission spectrum, the intensity was recorded with the wavelength step of 0.02nm. If the intensity at wavelength $\lambda$ was higher than those at $\lambda \pm d \lambda$, a peak/spike was identified at $\lambda$. Such method was called a three-point peak-finding. Using this method we found the spectral positions for the lasing peaks and the ASE spikes within the wavelength range 425-431nm from 200 single-shot emission spectra taken under identical conditions to obtain $P(\delta \lambda)$. Figure 3(a) shows the results obtained with the colloidal solution at three pumping levels. Below the lasing threshold, $P(\delta \lambda)$ decayed monotonically as $\delta \lambda$ increased from zero. It reflected the spacing statistics of ASE spikes, as the lasing peaks were absent. Note that the value of $P(\delta \lambda)$ at $\delta \lambda \sim 0$ could not be obtained, because two ASE spikes whose spacing was less than the spike width could not be resolved. Above the lasing threshold, a  local maximum of $P(\delta \lambda)$ was developed at $\delta \lambda \neq 0$. It shifted to larger $\delta \lambda$ as the pumping increased further. This local maximum originated from the regularity of lasing peak spacing. However its position was smaller than the average spacing obtained from $C(\Delta \lambda)$, because the three-point peak-finding method selected not only the lasing peaks but also the ASE spikes. 

Since the ASE spikes were narrower than the lasing peaks, we tried three different methods to unselect the ASE spikes. An ASE spike typically covered 3-4 $d \lambda$, while a lasing peak 7-8 $d \lambda$. We used the five-point peak-finding method, i.e., $I(\lambda) > I(\lambda \pm d \lambda) > I(\lambda \pm 2 d \lambda) $ to identify a peak at $\lambda$. The ASE spikes were narrow and closely packed, thus most of them were not selected. Consequently, $P(\delta \lambda)$ at $\delta \lambda$ close to 0 was greatly reduced. The local maximum at $\delta \lambda \neq 0$ shifted to larger $\delta \lambda$, and became the global maximum. An example is shown in the inset of Fig. 3(a) for $E_p$ = 0.13$\mu$J. To confirm this result, we tried a different method. The emission spectrum was first smoothed by the three-adjacent-point averaging, i.e., $I(\lambda)$ was replaced by $[I(\lambda-d \lambda)+I(\lambda)+I(\lambda+ d \lambda)]/3$. Most ASE spikes were smeared out by such process, while the lasing peaks survived. Then we used the three-point peak-finding method to find the peaks. The third method was a combination of the first two, a three-adjacent-point smoothing followed by a five-point peak-finding. As shown in the inset of Fig. 3(a), $P(\delta \lambda)$ obtained with all three methods were very close to each other. Especially $P(\delta \lambda)$ all reached the maximum at $\delta \lambda$ = 0.27nm, which coincided with the average lasing peak spacing obtained from the oscillation period of $C(\Delta \lambda)$. This result reflected the spectral repulsion of lasing modes. Since the particle suspension was in the diffusive regime, its quasimode spacing $\delta \lambda$ should satisfy the Wigner-Dyson distribution \cite{Beenakker_1997} $P(\delta \lambda) \sim \delta\lambda \exp[-\pi (\delta \lambda)^2/ 4 \langle \delta \lambda \rangle ^2]$. However, $P(\delta \lambda)$ for the lasing modes did not fit the Wigner-Dyson distribution well. The deviation might be caused by several factors such as the mode competition for gain which would limit the number of lasing modes \cite{cao_2003}, and the modification of lasing modes by local pumping even in the absence of reabsorption (to be discussed later). 

Figure 3(b) shows $P(\delta \lambda)$ for the ASE spikes from the neat dye solution at three pumping levels. The ASE spikes were selected by the three-point peak-finding method. The wavelength spacing statistics for the ASE spikes clearly differed from that for the lasing peaks. $P(\delta \lambda)$ always decayed monotonically with increasing $\delta \lambda$, even at high pumping. However the decay became slightly slower with increasing pumping. This change was caused mainly by the noise of the CCD array detector, which produced tiny spikes in the spectrum. Such noise spikes were also selected by our peak-finding program. However, they became less significant at higher pumping, as the ASE spikes dominated the emission spectrum. The inset of Fig. 3(b) is the log-linear plot of $P(\delta \lambda)$. With increasing pumping level $P(\delta \lambda)$ approached an exponential decay, as the ASE spikes dominated over the noise spikes. The solid line represents an exponential fit of the data at $E_p = 0.13 \mu$J, $P(\delta \lambda) \sim \exp(-\delta \lambda / 0.08)$. It suggested that the ASE spikes satisfied the Poisson statistics, which meant the frequencies of individual ASE spikes were uncorrelated. 

\section{Intensity statistics}

Further difference between the ASE spikes and lasing peaks was revealed in the intensity statistics. First we computed the average intensity at each wavelength $\langle I(\lambda) \rangle$ from 200 single-shot emission spectra taken under identical condition, and normalized the emission intensity $I(\lambda) / \langle I(\lambda) \rangle$. Then we extracted the statistical distribution of the normalized emission intensity within the wavelength range 425-431nm. Figure 4(a) is the log-log plot of $P(I/ \langle I \rangle)$ obtained with the colloidal solution at three pumping levels. Above the lasing threshold, $P(I/ \langle I \rangle)$ had a power-law decay at large $I$. Solid lines represent the fitting with $P(I/ \langle I \rangle) \sim (I/ \langle I \rangle)^{-b}$, where $b$ = 3.3, 2.5 for $E_p$ = 0.09$\mu$J, 0.13$\mu$J. The power-law decay became slower at higher pumping. Below the lasing threshold, e.g. at $E_p$ = 0.05$\mu$J, $P(I/ \langle I \rangle)$ did not fit the power-law decay well. It was similar to the intensity statistics of emission from the neat dye solution. Figure 4(b) is the log-linear plot of $P(I/ \langle I \rangle)$ obtained with the neat dye solution at the same pumping levels as those in Fig. 4(a). Instead of a power-law decay, $P(I/ \langle I \rangle)$ exhibited an exponential decay at large $I$, and the decay rate was nearly the same for different pumping. The solid line represents an exponential fit $P(I/ \langle I \rangle) \sim  \exp(-I/ \langle I \rangle/a)$ with $a$ = 0.2. Hence, the intensity statistics for the neat dye solution was very different from that for the colloidal solution above the lasing threshold. Such difference originated from the fact that the former had only ASE spikes and the latter had lasing peaks.  

Next we obtained the statistical distributions for the lasing peak height and ASE spike height. After normalizing 200 single-shot spectra from the colloidal solution under a constant pumping by their ensemble-average $\langle I(\lambda) \rangle$, we used the five-point peak-finding method described earlier to find the lasing peaks and record their heights $I_p$. The heights of all lasing peaks within the wavelength range 425-431nm were averaged to obtain $\langle I_p \rangle$. Figure 4(c) shows the statistical distribution $P(I_p / \langle I_p \rangle)$ for the colloidal solution at the same pumping levels as those in Fig. 4(a). For the two pumping levels above the lasing threshold, $P(I_p / \langle I_p \rangle)$ exhibited power-law decay at large $I_p$. The fitting with $P(I_p/ \langle I_p \rangle) \sim (I_p/ \langle I_p \rangle)^{-b}$ gave $b$ = 2.7, 2.2 for $E_p$ = 0.09$\mu$J, 0.13$\mu$J. The values of $b$ were slightly smaller than those for $P(I/ \langle I \rangle)$, indicating the tail of the distribution for the normalized lasing peak height was more extended than that for the emission intensity collected at every frequency. 
Similarly we obtained the statistical distribution for the ASE spike height $P(I_p/ \langle I_p \rangle)$ from the neat dye solution using the three-point peak-finding method. As shown in Fig. 4(d),  $P(I_p/ \langle I_p \rangle)$ exhibited an exponential decay at large $I_p$ for all three pumping levels, and the decay rate was nearly the same. The solid line represents the exponential fit for the data, $P(I_p/ \langle I_p \rangle) \sim \exp[I_p/   \langle I_p \rangle/a]$ with $a$ = 0.18. The value of $a$ was close to that for $P(I/ \langle I \rangle)$.      

The above experimental results of correlations and statistics demonstrated the fundamental difference between the ASE spikes and the lasing peaks. Next we present a qualitative explanation for the ASE spikes. The stochastic structures of the pulsed ASE spectra of neat dye solutions were observed long ago \cite{bor,szatmari,sperber}. In our experiment, tight focusing of the pump beam created a cone-shaped gain volume inside the neat dye solution. The amount of amplification that a spontaneously emitted photon experienced was determined by its path length inside the gain volume. Since photons in the neat dye solution experienced no scattering and traveled in straight line, their path length inside the excitation cone depended on where and in which direction they were emitted. The photons spontaneously emitted near the cone ends in the direction parallel to the cone axis experienced the largest amplification because of their longest path length inside the gain volume. The ASE at the frequencies of these photons was the strongest, leading to the spikes in the emission spectrum. Of course, these photons must be emitted in the beginning of the short pump duration, otherwise they would not pick up the transient gain. Although the spontaneous emission time was a few nanosecond, only the initial part of the spontaneous emission pulse was strongly amplified. The ASE pulse was a few tens of picoseconds long, followed by a spontaneous emission tail. The spectral width of the ASE spikes was determined by the ASE pulse duration. We extracted the average width of the ASE spikes from the width of spectral correlation function in Fig. 2(b). After taking into account the spectral resolution of our spectrometer, we estimated the ASE pulse duration to be around 20ps, in agreement with the pumping pulse duration. Since different ASE spikes originated from different spontaneous emission events which were independent of each other, their frequencies were uncorrelated. It led to the Poisson statistics of the frequency spacing of neighboring ASE spikes, $p(\delta \lambda) \sim \exp[- \delta \lambda / \langle \delta \lambda \rangle]$. 

In the neat dye solution, the intensity of an ASE spike could be written as 
\begin{equation}
I = I_0 e^{l/l_g},
\end{equation}
where $l$ is the path length inside the gain region, and $l_g$ is the gain length. If optical gain was uniform inside the excitation cone and constant in time, the distribution of path length inside the cone for spontaneously emitted photons $P(l)$ = constant. From Eq. (1), the statistical distribution of ASE intensity $P(I) \propto 1/ I$. Experimentally $P(I)$ decayed exponentially [Fig. 4]. This discrepancy was attributed to the spatial and temporal variations of optical gain, which made it difficult to calculate $P(I)$. 

Although the occurrence of ASE spikes did not rely on scattering, multiple scattering could increase the path lengths of spontaneously emitted photons inside the gain volume thus raise the amplitudes of some spikes. Note that Eq. (1) was no longer valid. In a scattering medium if a photon triggered the stimulated emission of a second photon, they propagated together till they reached the next scatterer by which they might be scattered into different directions and explore different paths afterwards. For those rare long paths, the photons generated along the way likely switched to shorter paths which were more probable. Thus the emission intensity at the end of a long path of length $l$ should be less than that given by Eq. (1). In our experiment, the tight focusing of the pump beam and the low particle density in the colloidal solution made the scattering mean free path exceed the size of gain volume. The effect of scattering on the photon path length inside the gain volume was negligibly small, thus the ASE spikes exhibited little dependence on the particle density. 

\section{Effect of local excitation on random lasing modes}

Before interpreting the data for the lasing peaks, we must understand how the lasing modes were formed with local pumping and negligible reabsorption. Due to the large aspect ratio of the excitation cone, the laser emission was highly directional, indicating lasing occurred along the cone. Such a quasi-1D system was approximated as an 1D random system in our numerical simulation. The aim of the numerical simulation was to gain physical insight of the lasing mode formation, instead of reproducing the experimental results. Recently we developed a numerical method based on the transfer matrix to calculate the quasimodes of 1D random systems as well as the lasing modes under global or local pumping \cite{wu_2007}. The boundary condition was that there were only outgoing waves through the system boundary. In a passive system such boundary condition gave the frequency and decay rate of every quasimode, while in an active system it determined the frequency and threshold gain of each lasing mode. Our method was valid for the linear gain up to the lasing threshold, with both gain saturation and mode competition for gain being neglected. Such simplification did not affect our goal of finding all the {\it potential} lasing modes regardless of the material-specific nonlinear gain. 

The 1D random system we simulated consisted of $N$ dielectric layers with refractive index $n_d$ in air. The layer thickness and layer-to-layer spacing were randomized. Optical gain was introduced as an imaginary part $n_i$ of the refractive index. In the presence of uniform gain (constant $n_i$ across the system), the lasing modes had one-to-one correspondence with the quasimodes. As long as the scattering was not too weak, the lasing modes were nearly identical to the quasimodes in both frequency and spatial profile \cite{wu_2007}. Experimentally the gain coefficient was not uniform but varied spatially, because the pump intensity decreased with the depth inside the solution. Due to the intense pumping, the absorption of dye molecules near the front window of the cuvette was saturated. Thus the gain coefficient would be constant within certain depth and then decay exponentially. To simulate such case, $n_i$ was set constant within length $L_1$ from one end of the system and followed by an exponential decay of length $L_2$. 

When the length of the pumped region $L_p = L_1 + L_2$ was much smaller than the total system length $L$, the number of lasing modes was found to be less than that of the quasimodes within the same frequency range, especially in a weakly scattering random system. For example, we calculated all the lasing modes within the wavelength range 500-750nm in random systems of $L$ = 24.1 $\mu$m and $n_d$ = 1.05. In the absence of gain or absorption, the localization length $\xi$ = 240$\mu$m. It was much larger than $L$, thus light transport was ballistic. We simulated ten random configurations with the same degree of disorder to obtain the average number of lasing modes $N_l$ for a fixed $L_p$. As $L_p$ decreased from 12$\mu$m to 4.8$\mu$m, $N_l$ was reduced from 22 to 17. If the entire system was excited uniformly, $N_l$ = 30, which was equal to the number of quasimodes of the passive system. The reduction of $N_l$ with $L_p$ was not a result of mode competition or gain saturation which was ignored in our calculation. It was neither due to reabsorption which could reduce the effective system size and the density of states \cite{yamilov_2005}, because absorption was not introduced to the structures. The only reason for the lasing mode number reduction was the local excitation. Since a local gain selectively amplified the feedback from the scatterers within the pumped region, its contribution to the lasing mode formation was greatly enhanced. Meanwhile, the feedback from the scatterers outside the pumped region was not amplified, and its contribution to lasing became relatively weak. Hence, the relative strengths of feedbacks from different parts of the random structure were changed by local pumping, so was the interference of these scattered waves. Consequently, some quasimodes failed to lase, and those did lase had different spatial profile (to be shown later). Nevertheless, the sublinear decrease of $N_l$ with $L_p$ reflected the fact that $N_l$ exceeded the number of quasimodes of the reduced random system defined by the pumped region (the unpumped part was removed). It indicated the feedback from the unpumped region was not negligible.  

The reduction of $L_p$ led to less number of lasing modes and larger mode spacing. This result was confirmed experimentally. By changing the stilbene 420 dye concentration, we varied the penetration depth of the pump beam which determined the length $L_p$ of excitation cone. The TiO$_2$ particle density was kept constant so that the random structures were effectively the same. The excitation cone was imaged through a side window of the cuvette using an objective lens and a CCD camera. Since laser emission was directed along the cone, only spontaneous emission was collected from the side. The spontaneous emission intensity reflected the local pump intensity, and its spatial distribution revealed the shape of the pumped region. 

From the oscillation period of $C(\Delta\lambda)$, we extracted the average lasing mode spacing $\langle \delta \lambda \rangle$. As shown in Fig. 5, $\langle \delta \lambda \rangle$ = 0.36nm and 0.19nm for $M$ = 12mM and 6mM, respectively. Comparing them to $\langle \delta \lambda \rangle$ = 0.27nm for $M$ = 8.5mM in Fig. 2(a), we concluded that the adjacent lasing mode spacing scaled with the dye concentration. This was due to the change of penetration depth $L_p$ of the pump light into the solution with different dye concentration. This was confirmed by the images of pumped regions through the side window of the cuvette shown in the insets of Fig. 5. Although the variation of $\langle \delta \lambda \rangle$ with $M$ was similar to our previous results in colloidal solutions with rhodamine 640 dye \cite{wu_2006}, the underlying mechanisms were different. Due to strong reabsorption by the rhodamine 640 molecules in the unpumped region, the emitted photons which left the pumped region has little chance of returning to it. Therefore, the feedback from the unpumped region was suppressed, and the lasing modes were spatially confined in the vicinity of the pumped region. In contrast, the reabsorption of stilbene 420 was extremely weak, and the lasing modes were spread over the entire solution. The reduction of lasing mode number and the increase of the lasing mode spacing were caused purely by the local amplification in our current experiment.  
    
The local pumping could also modify the spatial profile of a lasing mode drastically in a weakly scattering system. Figure 6 showed the intensity distribution $I_l(x)$ of lasing modes under local pumping and global pumping as well as the quasimode without pumping. For comparison, $I_l(x)$ was normalized such that the spatial integration over the entire random structure was equal to 1. With gain uniformly distributed across the entire random structure, the lasing mode profile (solid line) was almost the same as the quasimode of the passive system (dotted line). The envelop of $I_l(x)$ grew exponentially toward the system boundary due to the negative imaginary part of the wavevector $k_i$, with additional modulations caused by interference of scattered light \cite{wu_2007}. When gain existed only in part of the random structure, the lasing mode profile (dashed line) changed significantly. $I_l(x)$ no longer grew exponentially outside the pumped region where gain was zero. The relative weight of the intensity distribution in the part far from the gain region was reduced. Such redistribution was attributed to the change in the relative strength between the feedback from the pumped region and that from the unpumped region. 

Our numerical simulation reproduced the regular frequency spacing of lasing modes that was observed experimentally. We calculated the frequency spacing of adjacent lasing modes in random structures with various amount of scattering. As the scattering got weaker, the mode spacing became more regular. As an example, we calculated all the lasing modes within the wavelength range 500-750nm in a random structure (length $L$ = 24.1$\mu$m) under local pumping ($L_p$ = 9.6$\mu$m). Figure 7 plots the frequency spacing of adjacent lasing modes $\delta k_r$ normalized to the average value $\langle \delta k_r \rangle$ for $n_d$ = 1.05 (cross) and 2.0(square). The fluctuation of lasing mode spacing increased with $n_d$ or the scattering strength. This results agreed with the experimental observation that the frequency spacing of lasing peaks became less regular when we increased the amount of scattering by adding more particles to the solution. Our simulation illustrated that the regular spacing of lasing modes in a weakly scattering system was intrinsic, namely, it was not caused by mode competition for gain or gain saturation which was neglected in our calculation. As can be seen clearly by comparing Fig. 5(a) to (b), the degree of regularity in frequency spacing of lasing modes decreased as we increased $L_p$ by lowering the dye concentration. 

The deviation of the lasing modes under local excitation from the quasimodes might lead to  different statistical behaviors. We calculated the statistical distribution of frequency spacing $\delta k_r$ of quasimodes from 50 different random structures with $L$ = 24.1$\mu$m and $n_d$ = 1.05, as well as that of the lasing modes under local excitation $L_p$ = 12$\mu$m. Since the ensemble-averaged spacing $\langle \delta k_r \rangle$ of the quasimode might be smaller than that of the lasing modes, $\delta k_r$ was normalized by $\langle \delta k_r \rangle$ for comparison. As shown in Fig. 8(a), $P(\delta k_r / \langle \delta k_r \rangle)$ for the lasing modes was broader than that for the quasimodes, and it became asymmetric. We also calculated the statistical distribution of the decay rate $k_i$ of quasimodes, as well as that of the threshold of lasing modes under local pumping. Again $k_i$ was normalized by the ensemble-averaged value. Figure 8(b) illustrates that the two distributions were rather similar.

\section{Discussion and Summary}

Although the dimensions of the random systems in our numerical calculation were much smaller than the experimental ones, the calculation results provided qualitative explanation for the experimental data. Tight focusing of the pump light greatly reduced the number of potential lasing modes $N_l$. The increase of mode spacing facilitated the observation of discrete lasing peaks in the emission spectrum. Since the reduction of $N_l$ was a result of local amplification, it did not rely on the reabsorption outside the pumped region. The increase of $N_l$ with $L_p$ agreed qualitatively with the experimental observation of smaller lasing peak spacing for longer pump cone. One quantitative difference was that experimentally the samples were three-dimensional and the emitted photons which wandered out of the pumped region had less chance of returning to it via scattering. The contribution of the scatterers in the unpumped part to the lasing mode formation was weaker than that in the 1D random systems we calculated. Nevertheless, the lasing modes were extended over the entire random system, instead of being confined to the pumped region like the lasing modes in a system with strong reabsorption. The feedbacks from the scatterers outside the pumped region were not negligible, thus the lasing modes were not equal to the quasimodes of the reduced system defined by the gain volume. Our numerical calculation also showed that the frequency spacing of adjacent lasing modes under local pumping became more regular as the scattering strength decreased. Such regularity was intrinsic and not caused by mode competition for gain. Since local pumping could make the lasing modes dramatically different from those of the quasimodes of the passive system, the statistics of quasimodes \cite{nikolic_2001,kottos_2005} cannot be applied directly to predict the properties of lasing modes. To reproduce the statistical data of the lasing modes, one shall take into account mode competition and gain saturation which would make the calculation more complicated. In addition, the intensities of the lasing modes under pulsed pumping also depended on the initial spontaneous emission into individual modes, which varied from shot to shot. This led to shot-to-shot fluctuation of lasing peak height. Therefore, it was difficult to include all these factors in our current calculation, and we hope our work will stimulate further theoretical studies. 

In summary, we demonstrated experimentally the spectral correlation and intensity statistics for random lasing modes in a weakly scattering system were very different from those for the ASE spikes. Since the ASE spikes originated from independent spontaneous emission events, their frequencies were uncorrelated, leading to Poisson statistics for their spectral spacing.  The lasing peaks represented the lasing modes, which could be drastically different from the quasimodes due to local pumping. Our numerical simulation illustrated that even without reabsorption the number of potential lasing modes might be greatly reduced by local excitation of a weakly scattering system. The selective amplification of the feedback from the scatterers within the gain region could significantly modify the lasing mode profile. Our studies not only revealed the relation between the lasing modes and the quasimodes under global or local pumping, but also explained why tight focusing of the pump light facilitated the observation of discrete lasing peaks in the emission spectra.   

The authors acknowledge Profs. Alexey Yamilov and Andrey Chabanov for stimulating discussions. This work was supported by the National Science Foundation under Grant Nos. DMR-0093949 and ECS-0601249.

\newpage
\begin{figure}
\centerline{\scalebox{1.5}{\includegraphics{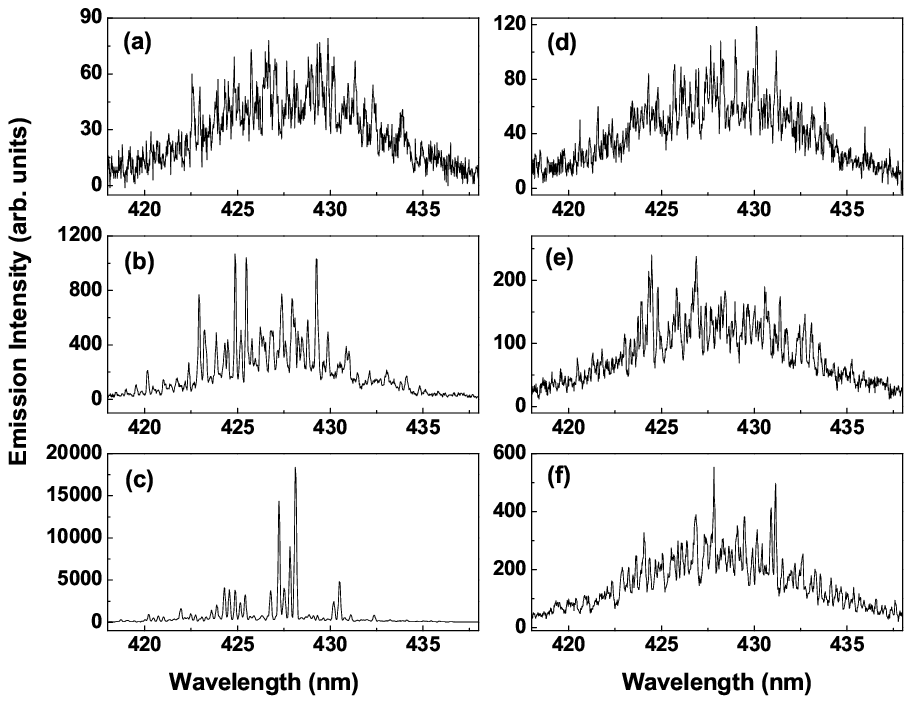}}}
\caption{Single-shot spectra of emission from the 8.5 mM stilbene 420 dye solutions with TiO$_2$ particle density $\rho$ = $3\times10^{9}$cm$^{-3}$ (a)-(c), and $0$ (d)-(f). The pump pulse energy $E_p$ = 0.05$\mu$J for (a) \& (d), 0.09$\mu$J for (b) \& (e), 0.13$\mu$J for (c) \& (f).}
\label{fig1}
\end{figure}
     
\newpage
\begin{figure}
%\centerline{\scalebox{0.8}{\includegraphics{fig2a.eps}}}
%\centerline{%\rotatebox{-90}{
%\begin{tabular}{c c}
\includegraphics[width=0.6\textwidth]{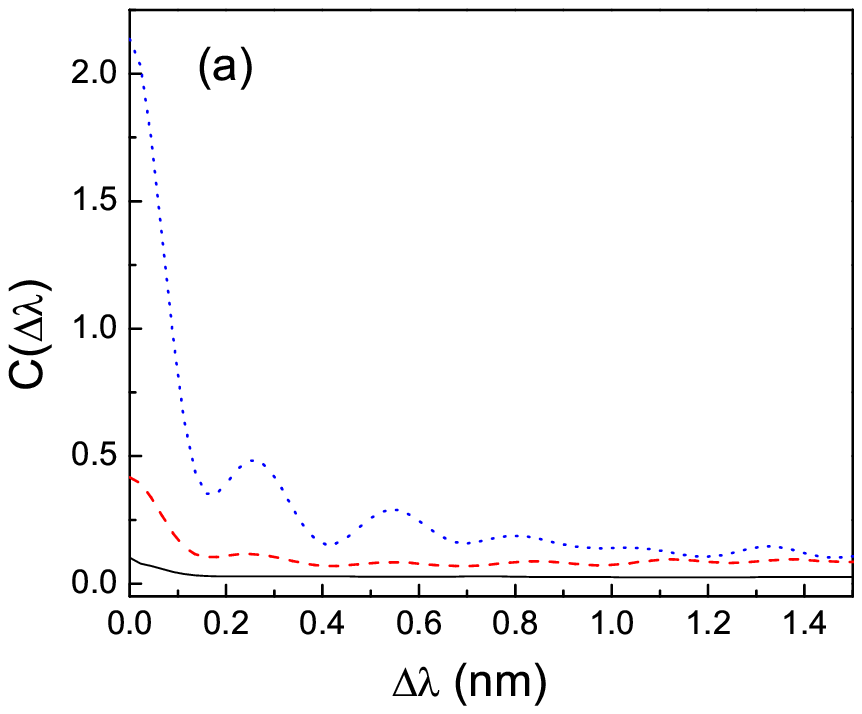} 
\includegraphics[width=0.6\textwidth]{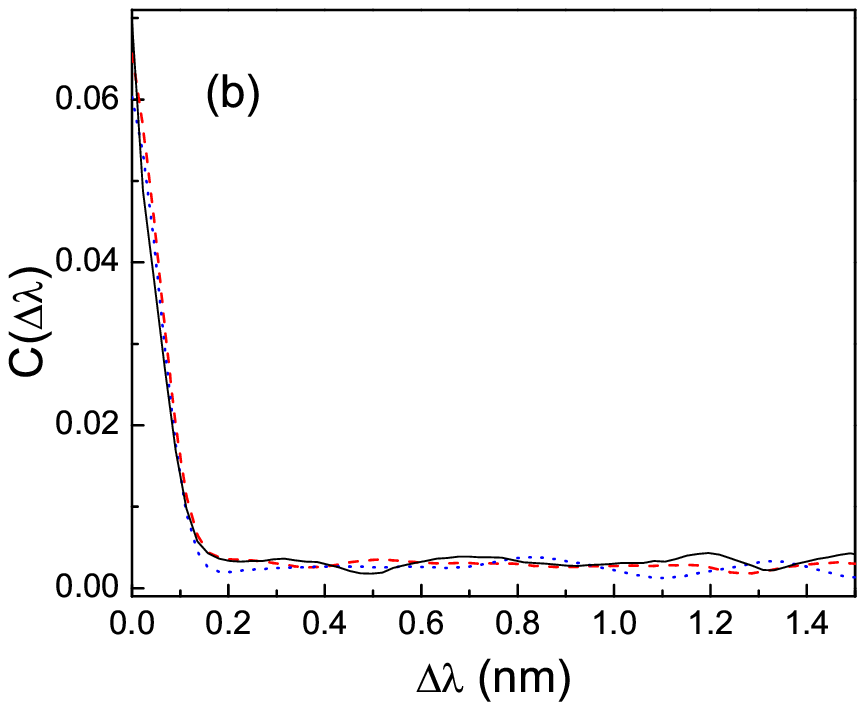}
%\end{tabular}}%}
\caption{Ensemble-averaged spectral correlation function $C(\Delta \lambda)$ of single-shot emission spectra. $M$ = 8.5 mM. $\rho$ = $3\times10^{9}$cm$^{-3}$ (a), $0$ (b). $E_p$ = 0.05$\mu{J}$(solid lines), 0.09$\mu{J}$(dashed lines) and 0.13$\mu{J}$(dotted lines).
}
\label{fig2}
\end{figure}

\newpage
\begin{figure}
%\centerline{%\rotatebox{-90}{
%\begin{tabular}{c c}
\includegraphics[width=0.5\textwidth]{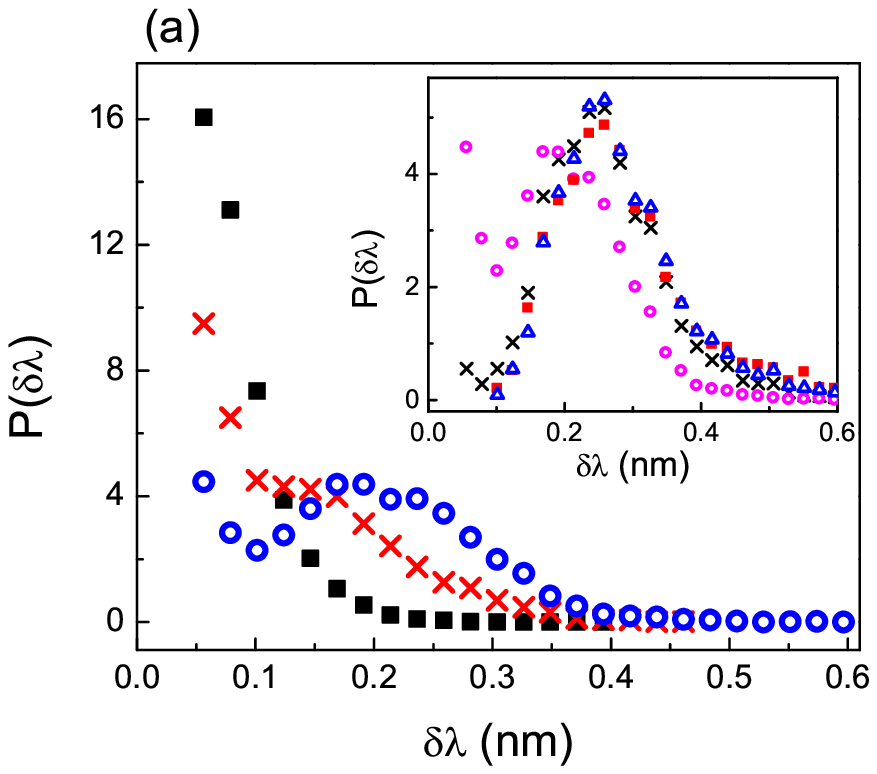} 
\includegraphics[width=0.5\textwidth]{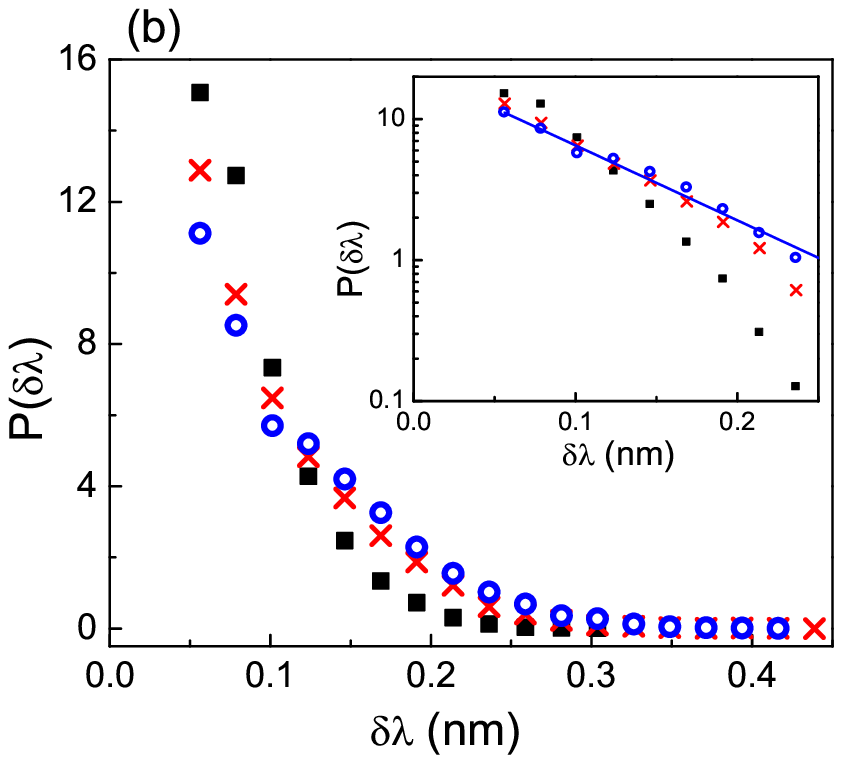}
%\end{tabular}}%}
%\centerline{\scalebox{0.8}{\includegraphics{fig3a.eps}}}
\caption{(Color online) Statistical distribution $P(\delta \lambda)$ of wavelength spacing $\delta \lambda$ between adjacent spikes/peaks selected by the 3-point peak-finding method. $\rho$ =  $3\times10^{9}$cm$^{-3}$ (a), $0$ (b). $E_p$ = = 0.05$\mu$J (squares), 0.09$\mu$J (crosses) and 0.13$\mu$J (circles). The inset of (a) shows $P(\delta \lambda)$ for $\rho = 3\times10^{9}$cm$^{-3}$ and $E_p$ = 0.13$\mu$J, obtained with four methods: 3-point peak-finding (circles) 5-point peak-finding (squares), 3-adjacent-point averaging followed by 3-point peak-finding (crosses) or 5-point peak-finding (triangles). The inset of (b) is the log-linear plot of $P(\delta \lambda)$ for $\rho = 0$, obtained with 3-point peak-finding method. $E_p$ = 0.05$\mu$J (squares), 0.09$\mu$J (crosses) and 0.13$\mu$J (circles). The solid line represents the exponential fit of the data for $E_p$ = 0.13$\mu$J: $P(\delta \lambda) = 22 \exp(- \delta \lambda / 0.08)$.}
\label{fig3}
\end{figure}

\newpage
\begin{figure}
\centerline{
\begin{tabular}{c c}
\includegraphics[width=0.45\textwidth]{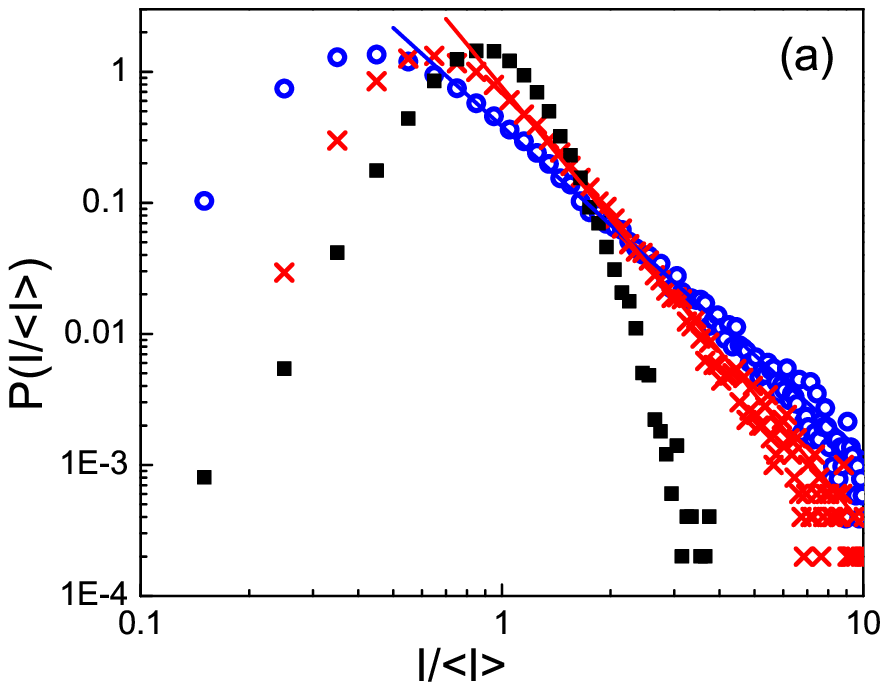} & \includegraphics[width=0.45\textwidth]{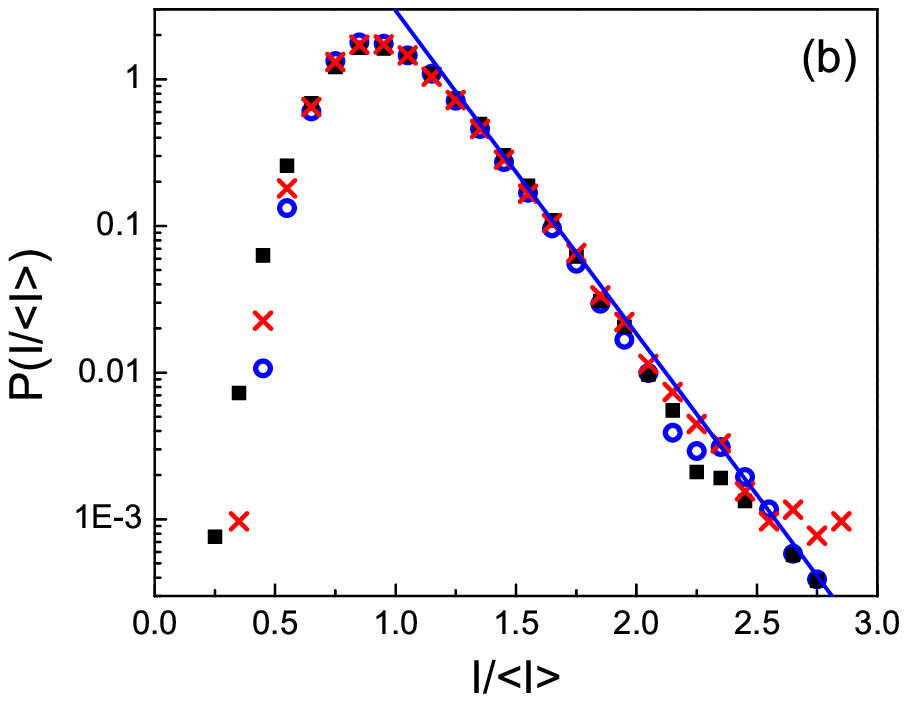}\\
\includegraphics[width=0.45\textwidth]{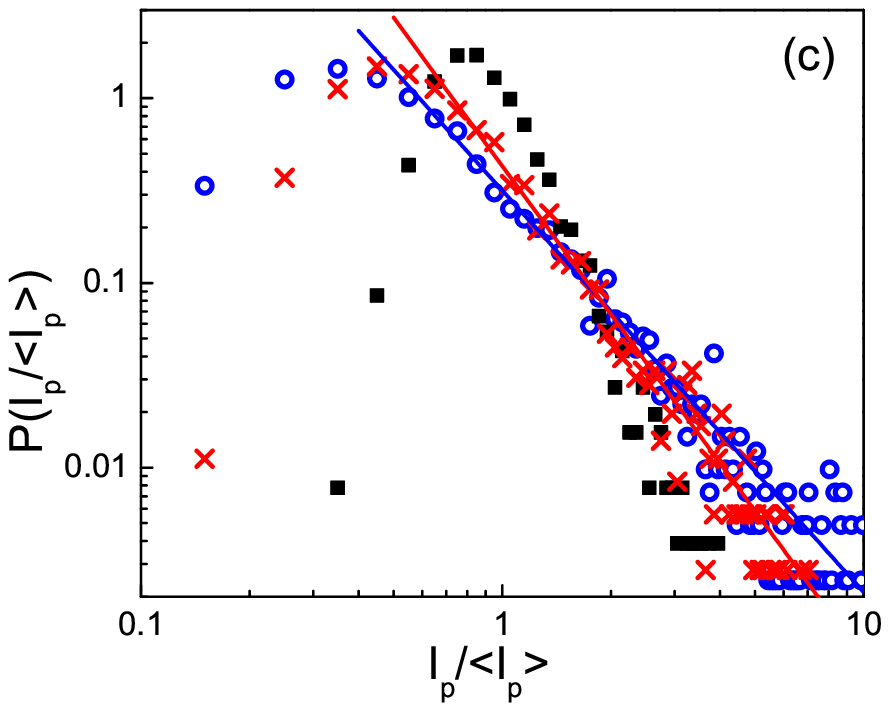} & \includegraphics[width=0.45\textwidth]{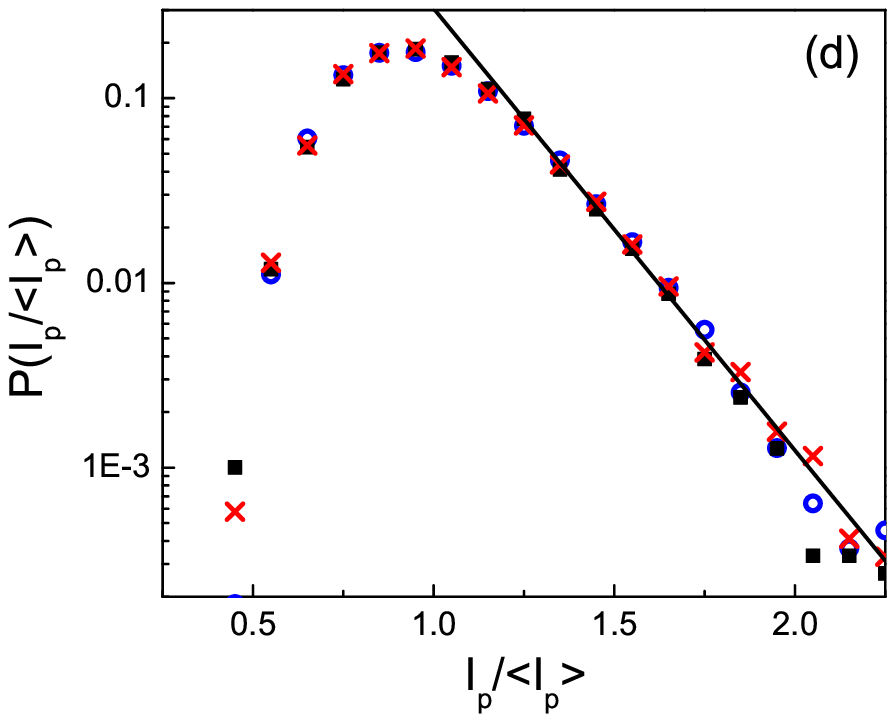}
\end{tabular}}
%\centerline{\scalebox{1.0}{\includegraphics{fig4.eps}}}
\caption{(Color online) Statistical distributions of normalized emission intensities $I(\lambda) / \langle I(\lambda) \rangle$ (a, b) and peak/spike heights $I_p / \langle I_p \rangle$ (c, d). $M$ = 8.5mM. $\rho$ = 0 (b, d) and $3 \times 10^{9}$cm$^{-3}$ (a, c). $E_p$ =  0.05$\mu$J (squares), 0.09$\mu$J (crosses) and 0.13$\mu$J (circles). The lasing peaks in (c) are selected by the 5-point peak-finding method. The ASE spikes in (d) are selected by the 3-point peak-finding method.  The solid lines represent the curve-fitting. In (a) $P(I / \langle I \rangle) = 0.767 (I / \langle I \rangle)^{-3.3}$ (for crosses) and $P(I / \langle I \rangle)=0.383 (I / \langle I \rangle)^{-2.5}$ (for circles). In (b), $P(I / \langle I \rangle) = 467 \exp(- I/ \langle I \rangle/0.2)$. In (c), $P(I_{p}/ \langle I_p \rangle)=0.43 (I_{p}/ \langle I_p \rangle)^{-2.7}$ (for crosses) and $P(I_p / \langle I_p \rangle)=0.32 (I_{p}/ \langle I_p \rangle)^{-2.2}$ (for circles). In (d), $P(I_p / \langle I_p \rangle)=76 \exp(-I_p / \langle I_p \rangle/0.18)$.} 
\label{fig4}
\end{figure}

\newpage
\begin{figure}
\centerline{\scalebox{0.6}{\includegraphics{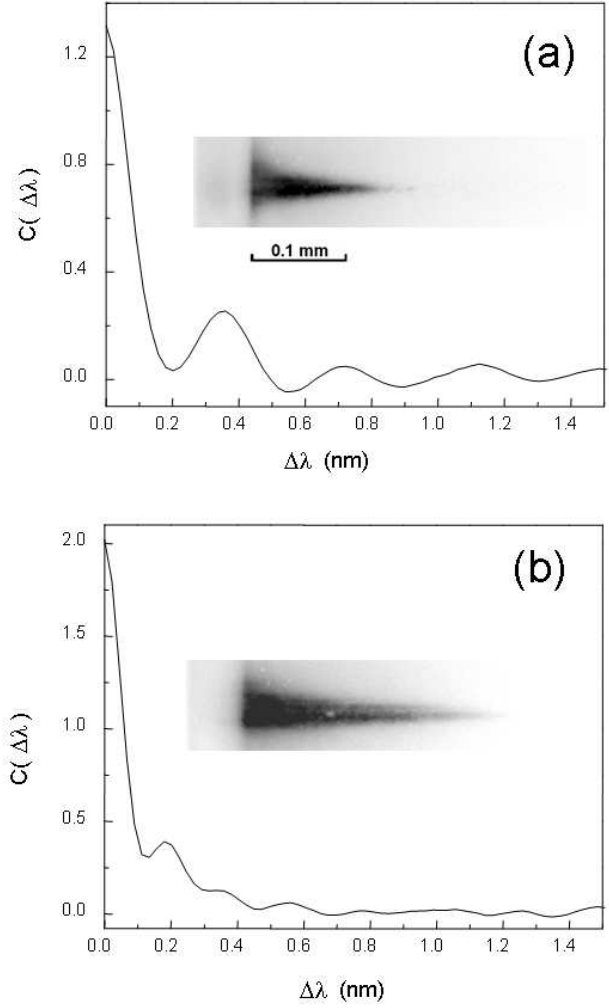}}}
\caption{Ensemble-averaged spectral correlation function $C(\Delta \lambda)$ of single-shot emission spectra from colloidal solutions with $\rho$ = $3 \times 10^{9}$cm$^{-3}$ TiO$_2$ particles and $M$ = 12mM (a) and 6mM (b). The insets are the images of the pumped region taken through the side window of the cuvette.} 
\label{fig5}
\end{figure}

\newpage
\begin{figure}
\centerline{\scalebox{1.0}{\includegraphics{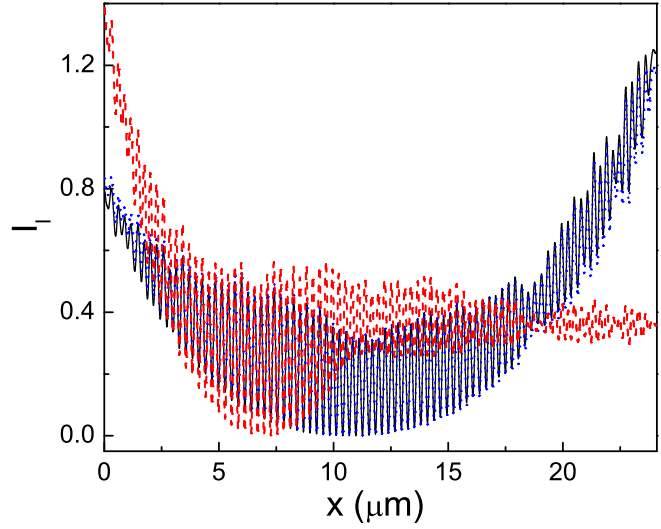}}}
\caption{(color online) Calculated spatial intensity distribution of a lasing mode at $\lambda$ = 575nm in an 1D random system under global pumping (black solid line), local pumping (red dashed line), and the corresponding quasimode of the passive system (blue dotted line). $L$ = 24.1$\mu$m, $n_d$=1.05, and $L_p$ = 9.6$\mu$m.}
\label{fig6}
\end{figure}

\newpage
\begin{figure}
\centerline{\scalebox{1.0}{\includegraphics{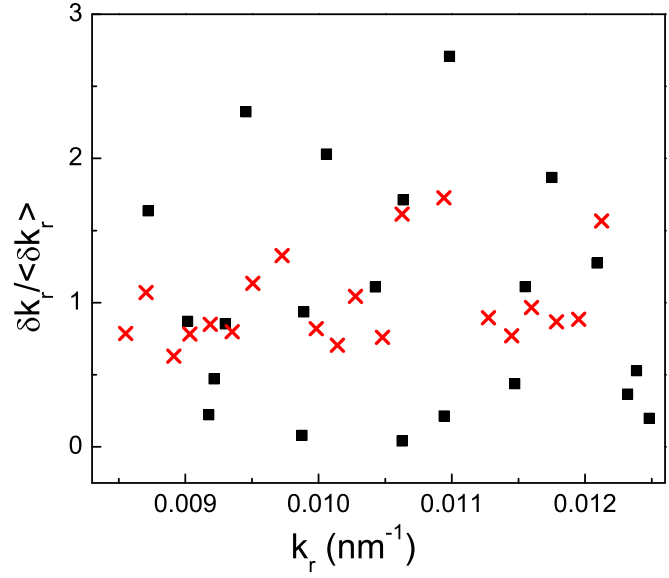}}}
\caption{Normalized frequency spacing $\delta k_{r}/\langle \delta k_{r} \rangle$ of neighboring lasing modes in 1D random systems with $n_d$ =2.0 (squares) and $n_d$ = 1.05  (crosses) under local excitation. $L$ = 24.1$\mu$m, and $L_p$ = 9.6$\mu$m.}
\label{fig7}
\end{figure}

\newpage
\begin{figure}
%\centerline{\scalebox{1.0}{\includegraphics{fig8.eps}}}
\includegraphics[width=0.6\textwidth]{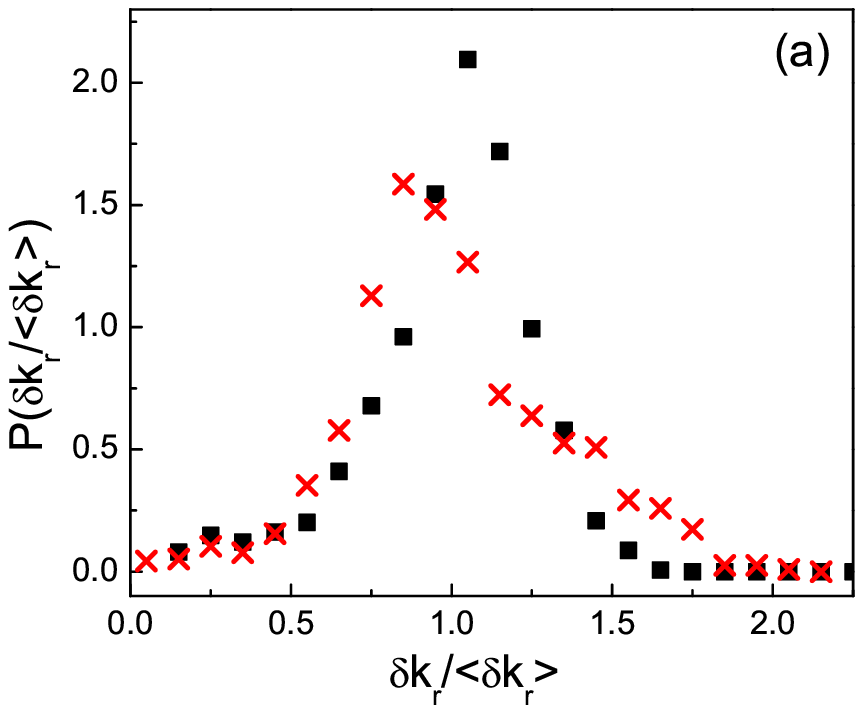} 
\includegraphics[width=0.6\textwidth]{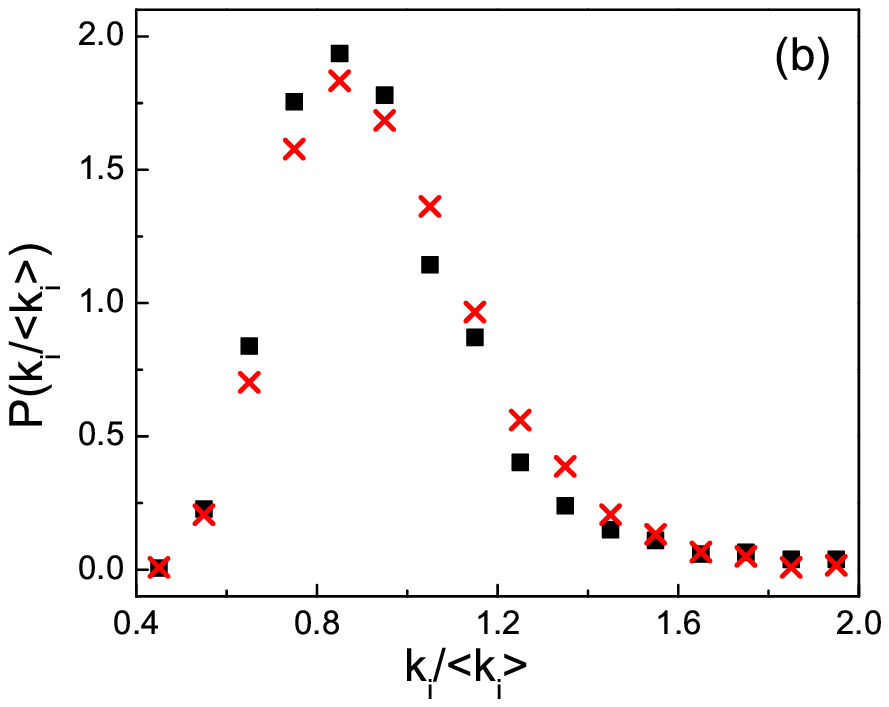}
\caption{(a) Statistical distribution $P(\delta k_r/\left\langle \delta k_r\right\rangle)$ of normalized frequency spacing between adjacent quasimodes (squares) and lasing modes under a local excitation (crosses). (b) Statistical distribution $P(k_i / \left\langle k_i \right\rangle)$ of normalized decay rates of quasimodes (squares) and the threshold gain of lasing modes under a local excitation (crosses). $L$ = 24.1$\mu$m, $n_d$=1.05, and $L_p$ = 12$\mu$m.}
\label{fig8}
\end{figure}


\begin{thebibliography}{}

\bibitem{cao_WRM} H. Cao, Waves in Random Media {\bf 13}, R1 (2003).

\bibitem{letokhov} R. V. Ambartsumyan, N. G. Basov, P. G. Kryukov,  and V. S. Letokhov, IEEE J. Quant. Electron.  {\bf QE-2} 442 (1966).

\bibitem{lawandy} N. M. Lawandy, R. M. Balachandran, A. S. L. Gomes, and E. Sauvain, Nature {\bf 368}, 436 (1994).

\bibitem{caoAPL98} H. Cao, Y. G. Zhao, H. C. Ong, S. T. Ho, J. Y. Dai, J. Y. Wu, and R. P. H. Chang, Appl. Phys. Lett. {\bf 73}, 3656 (1998).

\bibitem{caoPRL99}   H. Cao, Y. G. Zhao, H. C. Ong, S. T. Ho, E. W. Seelig, Q. H. Wang, and R.  P. H. Chang,  Phys. Rev. Lett. {\bf 82}, 2278  (1999).

\bibitem{vanneste_2001} C. Vanneste and P. Sebbah, Phys. Rev. Lett. {\bf 87}, 183903 (2001); 

\bibitem{jiang_2002} X. Y. Jiang, and C. M. Soukoulis, Phys. Rev. E {\bf 65}, 025601(R) (2002).

\bibitem{genack_2005}  V. Milner and A. Z. Genack,  Phys. Rev. Lett. {\bf 94}, 073901 (2005).

\bibitem{frolov_1999} S. V. Frolov, Z. V. Vardeny, K. Yoshino, A. A. Zakhidov, and R. H. Baughman, Phys. Rev. B {\bf 59}, R5284 (1999).

\bibitem{ling_2001} Y. Ling, H. Cao, A. L. Burin, M. A. Ratner, C. Liu, R. P. H. Chang, Phys. Rev. A {\bf 64}, 063808 (2001).

\bibitem{polson_2005} R. C. Polson, and Z. V. Vardeny, Phys. Rev. B {\bf 71} 045205 (2005).

\bibitem{wu_2006} X. Wu, W. Fang, A. Yamilov, A. A. Chabanov, A. A. Asatryan, L. C. Botten, and H. Cao, Phys. Rev. A, {\bf 74}, 053812 (2006).

\bibitem{apalkov_2002} V. M. Apalkov, M. E. Raikh, and B. Shapiro, Phys. Rev. Lett. {\bf 89}, 016802 (2002).

\bibitem{cao_2002} H. Cao, Y. Ling, J. Y. Xu, and A. L. Burin, Phys. Rev. E {\bf 66}, R25601 (2002).

\bibitem{vanneste_2007} C. Vanneste, P. Sebbah, and H. Cao, Phys. Rev. Lett. {\bf 98}, 143902 (2007).

\bibitem{burin_2001} A. L. Burin, M. A. Ratner, H. Cao, and R. P. H. Chang, Phys. Rev. Lett. {\bf 87}, 215503 (2001).

\bibitem{patra_2003} M. Patra, Phys. Rev. E, {\bf 67}, 016603 (2003).

\bibitem{hacken_2005} G. Hackenbroich, J. Phys. A: Math. Gen. {\bf 38}, 10537 (2005).

\bibitem{pinheiro_2006} F. A. Pinheiro and L. C. Sampaio, Phys. Rev. A {\bf 73} 013826 (2006).

\bibitem{lagendijk_2006} K. L. van der Molen, A. P. Mosk, and A. Lagendijk, Phys. Rev. A {\bf 74}, 053808 (2006).

\bibitem{deych_2005} L. I. Deych, Phys. Rev. Lett. {\bf 95}, 043902 (2005).

\bibitem{tureci_2006} H. E. T\"{u}reci, A. D. Stone, and B. Collier, Phys. Rev. A {\bf 74}, 043822 (2006).

\bibitem{tureci_2007} H. E. T\"{u}reci, A. D. Stone, and L. Ge, Phys. Rev. A {\bf 76}, 013813 (2007). 

\bibitem{yamilov_2005} A. Yamilov, X. Wu, H. Cao, and A. Burin, Opt. Lett. {\bf 30}, 2430 (2005).

\bibitem{mujumdar_2004} S. Mujumdar, M. Ricci, R. Torre, and D. W. Wiersma, Phys. Rev. Lett. {\bf 93}, 053903 (2004).

\bibitem{mujumdar_2007} S. Mujumdar, V. T\"{u}rck, R. Torre, and D. S. Wiersma, Phys. Rev. A {\bf 76}, 033807 (2007).

\bibitem{wu_2007a} X. Wu, and H. Cao, Opt. Lett. {\bf 32}, 3089 (2007). 

\bibitem{Beenakker_1997} C. W. J. Beenakker, Rev. Mod. Phys. {\bf 69}, 731 (1997).

\bibitem{cao_2003} H. Cao, X. Jiang, Y. Ling, J. Y. Xu, and C. M. Soukoulis, Phys. Rev. B. {\bf 67}, R161101 (2003).

\bibitem{bor} Z. Bor, S. Szatmari, and A. Muller, Appl. Phys. B {\bf 32}, 101 (1983).

\bibitem{szatmari} S. Szatmari and F. P. Sch\"{a}fer, Opt. Commun. {\bf 49}, 281 (1984).

\bibitem{sperber} P. Sperber, M. Weidner and A. Penzkofer, Appl. Phys. B {\bf 42}, 85 (1987).  

\bibitem{wu_2007} X. Wu, J. Andreasen, H. Cao, and A. Yamilov, J. Opt. Soc. Am. B {\bf 24}, A26 (2007).

\bibitem{nikolic_2001} B. K. Nikolic, Phys. Rev. B {\bf 65}, 012201 (2001). 

\bibitem{kottos_2005} T. Kottos, J. Phys. A: Math. Gen. {\bf 38}, 10761 (2005).

\end{thebibliography}
\end{document}